\documentclass[useAMS,usenatbib]{mn2e}
\usepackage{url,times,graphicx,amsmath,amsfonts,amssymb,aas_macros,color,epsfig,varioref,subfigure,comment,natbib,appendix}

\newcommand{\Tab}[1]{Table~\ref{#1}}
\newcommand{\Sec}[1]{Section~\ref{#1}}
\newcommand{\App}[1]{Appendix~\ref{#1}}
\newcommand{\Eq}[1]{Eq.~(\ref{#1})}
\newcommand{\Fig}[1]{Fig.~\ref{#1}}
\newcommand{\hMpc}{{\ifmmode{h^{-1}{\rm Mpc}}\else{$h^{-1}$Mpc}\fi}}
\newcommand{\hGpc}{{\ifmmode{h^{-1}{\rm Mpc}}\else{$h^{-1}$Gpc}\fi}}
\newcommand{\hkpc}{{\ifmmode{h^{-1}{\rm kpc}}\else{$h^{-1}$kpc}\fi}}

\newcommand{\hMsun}{{\ifmmode{h^{-1}{\rm {M_{\odot}}}}\else{$h^{-1}{\rm{M_{\odot}}}$}\fi}}
\newcommand{\Msun}{{\ifmmode{{\rm {M_{\odot}}}}\else{${\rm{M_{\odot}}}$}\fi}}
\def\hMpc{$h^{-1}\,{\rm Mpc}$}
\def\hkpc{$h^{-1}\,{\rm kpc}$}
\def\kms{{ \rm km} $s^{-1}$}
\def\LCDM{\ensuremath{\Lambda}CDM}

\def\vtan{$v_{\rm tan}$}
\def\vtanI{$v_{\rm tan}^{(I)}$}
\def\vtanII{$v_{\rm tan}^{(II)}$}
\def\vrad{$v_{\rm rad}$}
\def\std{Rand}

\title
[The tangential velocity of M31: CLUES from constrained simulations]
{The tangential velocity of M31: CLUES from constrained simulations} 
\author[Carlesi Edoardo]
{
Edoardo Carlesi, $^{1}$
\thanks{E-mail: carlesi@phys.huji.ac.il}
Yehuda Hoffman,$^{1}$
Jenny G. Sorce,$^{2}$
Stefan Gottl\"ober,$^{2}$\and
Gustavo Yepes,$^{5,6}$
H\'el\`ene Courtois,$^{3}$
R. Brent Tully$^{4}$
\\
\\
$^{1}$Racah Institute of Physics, Givat Ram, 91040 Jerusalem, Israel\\
$^{2}$Leibniz-Institut f\"ur Astrophysik Potsdam (AIP), An der Sternwarte 16, D-14482 Potsdam, Germany\\
$^{3}$University of Lyon, UCB Lyon 1/CNRS/IN2P3, F-69007 Lyon, France\\
$^{4}$Institute for Astronomy (IFA), University of Hawaii, 2680 Woodlaum Drive, HI 96822, US\\
$^{5}$Grupo de Astrof\'isica, Departamento de Fisica Teorica, Modulo C-8, Universidad Aut\'onoma de Madrid, E-280049 Cantoblanco, Spain\\
$^{6}$Astro-UAM, UAM, Unidad Asociada CSIC, E-280049 Cantoblanco, Spain\\
\setlength{\topmargin}{-1.5cm}
}
\begin{document}

\date{Submitted XXXX January XXXX}

\pagerange{\pageref{firstpage}--\pageref{lastpage}} \pubyear{2016}

\maketitle

\label{firstpage}

%%%%%%%%%%%%%%%%%%%%%%%%%%%%%%%%%%%%%%%%%%%%%%%%%%%

\begin{abstract}
Determining the precise value of the tangential component of the velocity of M31 is a non trivial astrophysical issue,
that relies on complicated modeling. 
This has recently lead to conflicting estimates, obtained by several groups that used
different methodologies and assumptions.
This \emph{letter} addresses the issue by computing a Bayesian posterior distribution function  of this quantity, in order to 
measure the compatibility of those estimates with \LCDM.
This is achieved using an ensemble of local group (LG) look-alikes collected from a set of Constrained Simulations (CSs) of the local Universe,
and a standard unconstrained \LCDM. 
The latter allows us to build a control sample of LG-like pairs and to single out the influence of the environment in our results.
%thanks to the faithful reproduction of our local neighbourhood embodied in the CSs sample.
We find that neither estimate is at odds with \LCDM; 
however, whereas CSs favour higher values of \vtan, the reverse is true for estimates based on LG samples gathered from unconstrained simulations, 
overlooking the environmental element.
\end{abstract}

%%%%%%%%%%%%%%%%%%%%%%%%%%%%%%%%%%%%%%%%%%%%%%%%%%%
%%%%%%%%%%%%%%%%%%%%%%%%%%%%%%%%%%%%%%%%%%%%%%%%%%%

\section{Introduction}\label{sec:intro}

The knowledge of the proper motion of M31 is required in order to constrain the properties and the evolution of the Local Group (LG).
Though the radial component of the velocity vector has been obtained more than one hundred years ago, 
the measure of its tangential velocity, \vtan,  is a much more challenging task.
In fact its value needs to be extrapolated 
through complicated modelling and under several hypotheses about the kinematics of satellite galaxies and its stellar populations.

A major breakthrough in this regard was represented by the work of \citet{Marel:2008} that was able to put an upper limit 
of $56$\kms\ to the transverse velocity of the M31 system from the line-of-sight motion of its satellites, 
assuming that they would on average follow their host galaxy.
%Modelling the kinematics of the different stellar populations of the system, 
\citet{Sohn:2012} were able to provide the first reconstruction of the M31 velocity vector itself, 
yielding a value of $17 \pm 17$\kms\ for its tangential component. This was done by measuring the displacement of 
different stellar populations within M31 with respect to reference galaxies in the background, using
some complicated modeling to single out the motion of the stars within M31 from the motion due to M31 itself.
However, these results have been recently challenged by \citet{Salomon:2016}, 
that using the precise measurements of satellite galaxy distances of \citet{Conn:2012}, 
obtain a value of \vtan$=164\pm61$\kms, at odds with the aforementioned analysis.
In the following, the estimates of \citet{Sohn:2012} and \citet{Salomon:2016} will be referred to as \vtanI\ and \vtanII\ respectively.

A different approach is followed here, which is based on a Bayesian inference of the value of \vtan. 
Such an inference starts with a prior knowledge of the system at hand and  a set of observations that are used 
to improve our knowledge of the tangential velocity of M31. 
This improved knowledge constitutes the posterior probability of \vtan\ given the prior knowledge and the observations. 
The inference of \vtan\ is based on the sampling of the posterior probability distribution function. 
The prior knowledge is split here into two independent models. 
One is the standard cosmological model that describes the Universe at large - and here the \LCDM\ model is assumed. 
The other is the LG model. Namely, it needs to be assumed a priori what is a LG. 
The  model can consist  of the basic dynamical characteristics of the LG, 
such as the distance and relative radial velocity of a pair of halos and their isolation. 
The model can be extended to include information about the merging history of the pairs of halos. 
The Cosmicflows-2 dataset of peculiar velocities \citep[CF2, ][]{Tully:2013} serves here as the observational data 
on our local 'patch' of the Universe. 
The  non-linear nature of the LG renders the analytical approach impossible and numerical simulations are to be used 
for sampling the likelihood and posterior probability distribution functions.
Constrained simulations are used to provide non-linear realizations of which obey both the prior \LCDM\ model and 
CF2 data \citep{Sorce:2015, Carlesi:2016a}. 
The ensemble of LG-like objects that emerge from the CF2 constrained simulations and the LG model 
provides a numerical sampling of the posterior probability. 
The ensemble  of LGs constructed from random simulations and the LG model sample the prior (likelihood) probability.

In this \emph{letter} we calculate the posterior probability of the  transverse velocity of the M31 galaxy (\vtan), 
assuming the \LCDM\ model and given the CF2 data. 
This work is structured as follows. The likelihood function of \vtan\ in a \LCDM\ universe is calculated as well and is compared with the posterior function.
\Sec{sec:methods}  describes the prior cosmological model, the  model used to define a LG and  the (constrained and random) simulations. 
The posterior distribution and the likelihood functions are presented in \Sec{sec:results}. 
A summary and an assessment of the implications of the results for the estimations of \vtan \citep{Sohn:2012,Salomon:2016} are given in \Sec{sec:end}.

%%%%%%%%%%%%%%%%%%%%%%%%%%%%%%%%%%%%%%%%%%%%%%%%%%%%%%%%%%%%%%%%%%%%%%%%%%%%%%%%%%

\section{Methods}\label{sec:methods}

\noindent
\textbf{Prior  cosmological model:} The prior model is assumed  here to be standard \LCDM\   with the Planck-I
cosmological   parameters of $\Omega _m = 0.31$, $\Omega _{\Lambda} = 0.69$, $h=0.67$ and $\sigma_8 = 0.83$ \citep{Planck:2013}.
A non-linear realization of the model is 
provided by the CurieHZ project \footnote{http://curiehz.ft.uam.es/} and consists in a (DM) only simulation done 
in a box size of $200$\hMpc\ with $1024^3$ particles, which will be referred to as \std\ hereafter.
DM halos are extracted from the simulation by the \texttt{AHF} halo finder \citep{Knollmann:2009} with halo mass  defined by $M_{200}$ (with respect
to $\rho_{c}$). The choice of this mass definition has only but a minor impact on \vtan\ as we checked by comparing to the results obtained with $M_{vir}$, 
the mass at the viral radius.

\noindent
\textbf{Prior LG model:}
Some of the fairly indisputable observational facts that describe the LG are the distance and relative radial velocity (\vrad)  
between the MW and M31 galaxies and the lack of a third comparable companion within a distance of 3 Mpc .
This leads us to formulate a very simplified prior model of the LG. 
An isolated pair of halos separated by a distance of $(0.35  $ -- $  0.70)$\hMpc\ and with \vrad\ in the range of $(-135$  --   $-80)$\kms, 
is defined be a LG like object. 
Isolation is defined by the absence of a halo more massive than the least massive of the two within $2.5$\hMpc. 
The ranges of  \vrad\ and $r$ are within $\pm 25\%$ of the fiducial values taken from \citet{Marel:2012}.
The prior model reflects the prior knowledge, or sometime prejudice, one has on the system at hand. 
The mass of the LG is far from being precisely known and estimates of it ranges over a factor of a few. 
Here we are willing to entertain the idea of introducing a mass range into the LG model. 
In the following we shall use two LG models, or priors; 
one with no mass constraints and one in which mass of the LG (defined as the sum of the $M_{200}$ of the two main haloes)
ranges over $(0.5$ -- $5.0)$\hMsun (see \Tab{tab:priors}).
That mass range reflects the uncertainties in various attempts to estimate the mass of the LG 
\citep[see e.g.][]{Klypin:2002, Widrow:2005, LiWhite:2008, Karachentsev:2009, Marel:2012}.

\begin{table}
\begin{center}
\caption{Kinematic priors on radial velocities (in \kms\ units) and relative distances of the haloes (in \hMpc); mass priors are given in
$10^{12}$\hMsun\ units. $N_{CS}$ and $N_{\std}$ are the numbers of halo pairs satisfying the criteria in the samples drawn from CS and \std\ simulations.}
\label{tab:priors}
\begin{tabular}{ccccc}
\hline
$V_{rad}$ & $r$ & $M_{LG}$ & $N_{CS}$ & $N_{\std}$ \\
\hline
$[-135, -80]$ & $[0.35, 0.70]$ & - & $1004$ & $857$ \\
$[-135, -80]$ & $[0.35, 0.70]$ & $[0.5, 5.0]$ & $173$ & $365$ \\
\hline
\end{tabular}
\end{center}
\end{table}

\noindent
\textbf{CF2 data and Constrained simulations:}
The bulk of the work described here relies on a set of 500 DM-only zoom simulations generated using the Local Group Factory 
\citep{Carlesi:2016a}. The Local Group Factory is a numerical pipeline designed for the production of $N$-body zoom simulations 
of LG-like objects within a large scale environment (Virgo, filament and Local Void) that closely matches the observational one.
The initial conditions for these simulations are generated using peculiar velocity data (taken from the CF2 catalog of \citet{Tully:2014}) as constraints;
for a detailed description of the physical and mathematical aspects of the CS technique we refer to the works of 
\citet{Doumler:2013a, Doumler:2013b, Doumler:2013c} and \citet{Sorce:2014, Sorce:2016}.
The cosmology used is of the Planck-I type, as in the case of the \std\ simulation, 
whereas the particle mass in the high resolution region is $m_p = 6.57\times10^8$\hMsun.
Our second sample (labeled CS) can straightforwardly obtained applying the aforementioned LG prior model to our set of CS simulations, 
ensuring that all of these objects live within an environment whose main features are remarkably close to the actual one. 

\noindent
\textbf{Sampling}
The sampling  of the posterior distribution  and the  likelihood functions of \vtan\ of the M31 follows the procedure used by 
\citet{Busha:2011}  and  \citep{Gonzalez:2014} for the estimation of the mass of the MW and the LG, respectively. 
We select ensembles of pairs of halos which obey the LG model from CF2-constrained and from random \LCDM\ simulations;
these samples provide numerical realizations of the  posterior probability and the likelihood functions. 
In the case of the likelihood function the search for the LGs extends over the full computational box, 
where in the posterior case the search is conducted within a sphere  of $7$\hMpc\ around the box center for each realization.
These realizations are used to calculate the mean and scatter of \vtan\  and to provide analytical fits to the above distribution function.
The sampling of the posterior distribution  and the  likelihood functions is performed twice - without and with the LG mass prior.

\section{Prior and posterior probability of the tangential velocity}\label{sec:results}

\begin{figure*}
\begin{center}
$\begin{array}{cc}
\includegraphics[height=7.0cm]{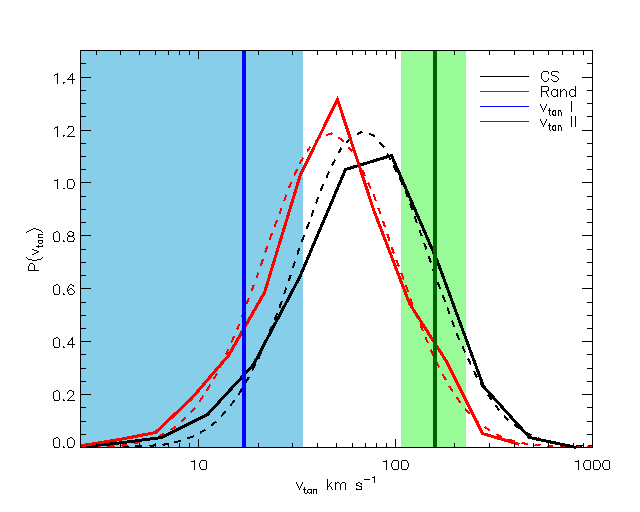} & 
\includegraphics[height=7.0cm]{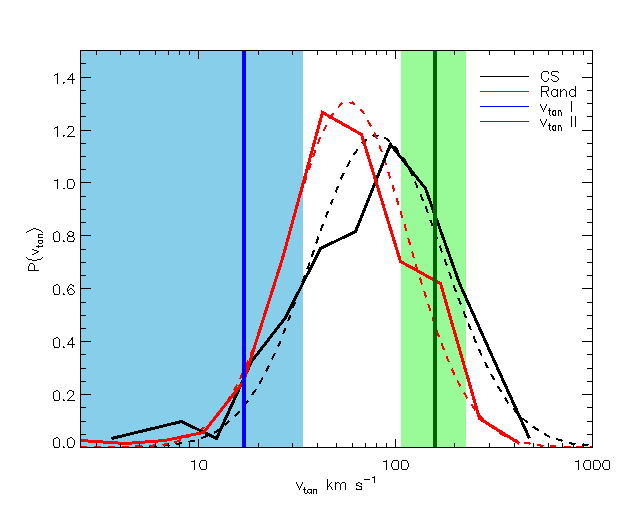} 
\end{array}$
\caption{
\label{fig:bayes}
\vtan\ distributions (thick lines) and fits to a log-normal distribution (dashed lines) for the CS and \std\ samples. 
Left panel: \Tab{tab:priors}, with no $M_{LG}$ prior. Right panel: same priors with $M_{LG}=[0.5, 5]\times10^{12}$\hMsun.
The vertical blue line stands for the \vtanI\ value while the green one for \vtanII; the shaded areas around them
indicate the 1$\sigma$ intervals.}
\end{center}
\end{figure*}

The four variants of the posterior probability  introduced in the previous paragraph are shown in \Fig{fig:bayes}.
Taking the $\log_{10}$ of \vtan\ it is possible to see that these distributions are well matched by a Gaussian function:

\begin{equation}\label{eq:gauss}
f(\log_{10}v) = \frac{1}{\sqrt{2\pi}\sigma} \exp{-\frac{(\log_{10}v - \mu)^2}{2\sigma^2}}
\end{equation}

\noindent
whose resulting means and standard deviations, for both the numerical and analytical best fit values, are shown in \Tab{tab:vtan}.

\begin{table}
\begin{center}
\caption{Mean and standard deviations of the $\log_{10}($\vtan$)$ distribution. The table contains numerical ($\mu_N$ and $\sigma_N$) and
best fit values to a Gaussian function ($\mu$ and $\sigma$), a different set of priors on radial velocity and relative distance.}
\label{tab:vtan}
\begin{tabular}{cccccc}
\hline
Sample & $\mu _N$ & $\sigma _N$ & $\mu$ & $\sigma$ & $M$ Prior\\
\hline
 CS &    1.97 &    0.36 &    1.89 &    0.34  & NO \\
 Rand &    1.77 &    0.35 &    1.71 &    0.33 & NO \\
 CS &    1.97 &    0.36 &    1.89 &    0.33  & YES\\
 Rand &    1.77 &    0.35 &    1.72 &    0.32 & YES \\
\hline
\end{tabular} \\
\end{center}
\end{table}

In general, one can test the sensitivity of the $P($\vtan$)$ to the choice of the interval, repeating the above computation using
larger (narrower) selection intervals on \vrad\ and $r$. 
This is true also when adding mass constraints, that have the only effect of substantially reducing the 
size of the sample, without altering the properties of the posterior distribution function. The results of these tests, which are presented in \App{app:priors},
confirm that the distributions are at best weakly sensitive on the priors, similarly to what \citet{Gonzalez:2014} have found in the case of 
the $P(M_{LG})$.

\begin{table}
\begin{center}
\caption{Fiducial values with $\pm \sigma$ upper and lower values for \vtanI, \vtanII\, and theoretical best-fit value estimates from the
CS and \std\ simulations (analytical estimate, no mass priors), in units of \kms.}
\label{tab:vtans}\begin{tabular}{cccc}
\hline
$\quad$ & $v$ & $v + \sigma _v$ & $v - \sigma_v$\\
\hline
\vtanI		& 17 	& 34 	& 0 \\
\vtanII		& 164 	& 225 	& 103 \\
\vtan$^{CS}$ &      78 &     168 &      36 \\
\vtan$^{Rand}$ &      51 &     109 &      24 \\
\hline
\end{tabular}
\end{center}
\end{table}

These posterior distribution functions yield the theoretical \vtan s, which are compared to the observational ones in \Tab{tab:vtans},
where it is shown that both \vtan s are in substantial agreement with \LCDM\ and \LCDM+CF2 predictions.
In fact, the overlap between the 2$\sigma$ intervals is nonzero, even though the peaks of $P($\vtan$)$ for the two samples tend to be
far from both \vtanI\ and \vtanII.
To calculate the compatibility of the two estimates with the theoretical predictions, that is their posterior probability or
degree of belief, we compute the integral:

\begin{equation}\label{eq:prob}
\int _{v_0 - \Delta v}^{v_0 + \Delta v} d v^{\prime} f(v^{\prime})
\end{equation}

\noindent
where $f(v^{\prime})$ is the Gaussian of \Eq{eq:gauss}, $\sigma$ and $\mu$ are the best fit values of \Tab{tab:vtan} and $v_0$ is given by the $\log_{10}$ 
values of \vtanI\ and \vtanII. The intervals $\Delta v$ are chosen to be the 1$\sigma$ values relative to \vtanI\ and \vtanII\ (see \Tab{tab:vtans}).
The numerical equivalent of the integral is simply given by

\begin{equation}\label{eq:p_num}
P^{(N)}_{I, II} = \frac{N_{v\pm\Delta v}}{N_{tot}}
\end{equation}

\noindent
where $N_{tot}$ is the total number of LG pairs of the sample and $N_{v\pm\Delta v}$ is the number of haloes within the intervals around the fiducial \vtan s.

The results of these calculations are shown in \Tab{tab:vtan2}, showing that $P_I > P_{II}$ for the \std\ simulations and $P_I < P_{II}$ in the case
of the CS set, consistently for both numerical and analytical estimates.
Moreover, the ratio of the two probabilities indicates that whereas \vtanI\ would be favoured by \emph{vanilla} \LCDM, with $\frac{P_I}{P_{II}}>1.27$, 
the opposite is true when adding environmental constraints (as given by the CF2 data). In the latter case, in fact, the probabilities are reversed and
the ratio $\frac{P_I}{P_{II}} < 0.5$ indicates that \vtanII\ kind of velocities are favoured when considering a LG pairs within 
a more realistic reconstruction of the Universe.

\begin{table}
\begin{center}
\caption{Probabilities of \vtanI\ and \vtanII\ for CS and \std\ LG samples and the different prior sets.
$P^A_I, P^A_{II}$ refer to the analytical estimates computed via \Eq{eq:prob}, while
$P^N_I, P^N_{II}$ are the numerical values, derived as the fractions of pairs within those intervals.
Remarkably, these latter values are completely unchanged by the restriction of $M_{LG}$ values, despite the large shrink in the 
halo sample size.} 
\label{tab:vtan2}
\begin{tabular}{cccccc}
\hline
Sample & $P_{I}^A$ & $P_{II}^A$ & $P_I^N$ & $P_{II}^N$& $M_{LG}$ Prior \\
\hline
 CS &    0.14 &    0.28 &    0.12 &    0.34  & NO \\
 Rand &    0.29 &    0.16 &    0.23 &    0.18 & NO \\
 CS &    0.14 &    0.29 &    0.12 &    0.34  & YES \\
 Rand &    0.27 &    0.16 &    0.23 &    0.18 & YES \\
\hline
\end{tabular}
\end{center}
\end{table}

%%%%%%%%%%%%%%%%%%%%%%%%%%%%%%%%%%%%%%%%%%%%%%%%%%%%%%%%%%%%%%%%%%%%%%%%%%%%%%%%%%

\section{Conclusions}\label{sec:end}

This work provides the posterior probability of the tangential velocity (\vtan) of the M31 galaxy assuming the \LCDM\ standard model of cosmology, 
a LG model which defines what is a LG, and given the Cosmicflows-2 database of peculiar velocities. 
This is compared with the likelihood function of \vtan\ given the prior \LCDM\ and the LG models. 
The sampling of the posterior probability and the likelihood function  is done by extracting all   LG-like objects 
from an ensemble of CF2 constrained \LCDM\  simulations via the so-called Local Group Factory \citep{Carlesi:2016a} 
and from a random \LCDM\ simulation, respectively.  
A simple LG model which specifies the distance, relative radial velocity  and the degree of isolation of a pair of halos is used to extract the LG-like objects. 
We used numerical and best-fit analytical expressions to derive \emph{predictions} of \LCDM\ model (and \LCDM+CF2 data, in the case of CS) for \vtan.
Posterior probabilities for \vtanI\ and \vtanII (defined as the \vtan\ obtained by \citet{Sohn:2012} and \citet{Salomon:2016}) were also computed, 
integrating the distributions over the 1$\sigma$ \vtan\ intervals of the two observations.

The main findings can be summarized as follows:

\begin{itemize}
\item The posterior probability and the likelihood function are both well approximated by a lognormal distribution (with respect to $\log$\vtan). 
The posterior probability and the likelihood function are quite similar, with a small offset of their parameters.

\item The mean and standard deviations for \vtan\ are  $78^{+90}_{-42}$\kms\ for the posterior distribution (i.e. CF2-constrained simulations, CS)  
and $51^{+58}_{-27}$\kms\ for prior \LCDM\ model (i.e. the \std\ simulation).

\item Both  \vtanI\ and \vtanII\ estimates are in agreement with the posterior distribution and the prior \LCDM\ model to within the 2$\sigma$ compatibility range.
However, it was consistently found that \LCDM+CF2 tends to favour \vtanII\ over \vtanI\, whereas the reverse is true for \LCDM-only based estimates. 

\item The present results are largely insensitive to the choice of priors.  
We have shown that both the posterior probability and the likelihood function are very weakly affected by 
modification of the LG-samples induced by altering the LG model used here. 
This includes a very small sensitivity to the assumed  mass range of the LG. 
This result is akin to the one found by \citet{Gonzalez:2014}, using a similar approach for the MW mass.
\end{itemize}

So, while both \vtanI\ and \vtanII\ values are possible outcomes of prior \LCDM\ and the posterior \LCDM + CF2, 
using constrained and random simulations we showed that 
the peculiar nature of the LG environment, as manifested by the CF2 data, mildly favours a higher \vtan, 
preferring the \citet{Salomon:2016} result over the estimate of \citet{Sohn:2012}.
We conclude by adding that the approach used here, that relies heavily on the use of large samples of CSs produced with the 
so called Local Group factory, can be extended to analyze and derive posterior distributions for many other LG properties,
such as e.g. satellite populations and mass, which are the subject of current investigation.

\section*{Acknowledgements}

EC would like to thank the Lady Davis Fellowship Fund for financial support. 
JS acknowledges support from the Alexander von Humboldt foundation.
YH has been partially supported by the Israel Science Foundation (1013/12).
SG and YH acknowledge support from DFG under the grant GO563/21-1.
GY thanks MINECO (Spain) for financial support under  project grants AYA2012-31101 and AYA2015-63810-P.
We thank the anonymous referee for his useful remarks.
We also thank the Red Espa\~nola de Supercomputaci\' on   for  granting us 
computing time  in the Marenostrum Supercomputer at the BSC-CNS where 
part of the analyses presented in this paper  have been performed, as well as PRACE for granting computing time in 
the CURIE supercomputer  where the CurieHZ simulations was done.

%%%%%%%%%%%%%%%%%%%%%%%%%%%%%%%%%%%%%%%%%%%%%%%%%%%

\bibliographystyle{mn2e}
\bibliography{../biblio}

\bsp

%%%%%%%%%%%%%%%%%%%%%%%%%%%%%%%%%%%%%%%%%%%%%%%%%%%

\appendix

\section{Local Group model and posterior probability}
\label{app:priors}

In principle, the properties of the posterior probability derived in this work might be dependent on the specific
choice of our prior Local Group model, for which many different prescriptions are allowed.
The trade-off between sample size and closeness to the observations
(in particular with regard to $r$ and \vrad, which are strongly constrained by the data)
has lead to a choice of intervals of $\pm 25\%$ around the fiducial values.

We will now show that the main results of the paper are largely unaffected by this specific choice of the priors, by repeating
the procedures explained in \Sec{sec:results} with the new LG-like pairs selected according to the new criteria.
These are shown in \Tab{tab:priorsA}, where the new values for \vrad\ and $r$ are now taken to be $\pm 2\sigma$ 
and $\pm 50\%$ around the fiducial values; with and without additional restrictions on the mass.
With respect to the results of \Tab{tab:vtan}, the best fit and numerical $\mu$s and $\sigma$s undergo very 
small changes, and the posterior probability derived using the CS sample keeps peaking at larger values than the one computed 
using the \std\ LG-like haloes.
The most important change that can be noticed is the reduction in the standard deviations of the 2$\sigma$ samples, for the Rand simulations
alone. This can be shown to be related to the final mass distribution of the LGs for this choice of the priors.
However, this correlation is just mentioned here as its implications are outside
the scope of this work, as we plan to discuss it in depth in an upcoming paper.

These findings for the distributions imply that the numerical and analytical probabilities computed with them are expected to give very similar results.
This is indeed the case, as can be seen by looking at \Tab{tab:vtanA2}, where these numbers are calculated explicitly.
We have therefore shown that our results are not biased by the choice of the prior as both more and less restrictive choices
lead to the same results.\\

\begin{table}
\begin{center}
\caption{
Alternative priors on \vrad\ (in \kms\ units), $r$ (in \hMpc) and masses (in $10^{12}$\hMsun\ units). 
$N_{CS}$ and $N_{\std}$ are the numbers of halo pairs for each prior set, for the CS and \std\ simulations.
}
\label{tab:priorsA}
\begin{tabular}{cccccc}
\hline
Prior & \vrad\ & $r$ & $M_{LG}$ & $N_{CS}$ & $N_{\std}$ \\
\hline
Pr.1 & $[-120, -100]$ & $[0.45, 0.57]$ & - & $118$ & $113$ \\
Pr.2 & $[-165, -55]$ & $[0.25, 0.78]$ & - & $1806$ & $4021$ \\
Pr.3 & $[-120, -100]$ & $[0.45, 0.57]$ & $[0.5, 5.0]$ & $22$ & $55$ \\
Pr.4 & $[-165, -55]$ & $[0.25, 0.78]$ & $[0.5, 5.0]$ & $259$ & $1599$ \\
Pr.5 & $[-165, -55]$ & $[0.25, 0.78]$ & $[1.0, 3.0]$ & $184$ & $1167$ \\
\hline
\end{tabular}
\end{center}
\end{table}

\begin{table}
\begin{center}
\caption{Numerical and best-fit values for the standard deviation $\sigma$ and mean $\mu$, for the four
sets of priors of \Tab{tab:priorsA}.}
\label{tab:vtanA1}
\begin{tabular}{ccccc}
\hline
Sample & $\mu _N$ & $\sigma _N$ & $\mu$ & $\sigma$ \\
\hline
Pr.1, CS &    1.99 &    0.37 &    1.88 &    0.35 \\
Pr.1, Rand &    1.80 &    0.28 &    1.72 &    0.24 \\
\hline
Pr.2, CS &    1.94 &    0.37 &    1.89 &    0.34 \\
Pr.2, Rand &    1.72 &    0.40 &    1.69 &    0.38 \\
\hline
Pr.3, CS &    1.99 &    0.37 &    1.82 &    0.35 \\
Pr.3, Rand &    1.80 &    0.28 &    1.72 &    0.23 \\
\hline
Pr.4, CS &    1.94 &    0.37 &    1.88 &    0.33 \\
Pr.4, Rand &    1.72 &    0.40 &    1.68 &    0.38 \\
\hline
Pr.5, CS &    1.90 &    0.36 &    1.82 &    0.30 \\
Pr.5, Rand &    1.78 &    0.34 &    1.69 &    0.29 \\
\hline

\end{tabular}

\caption{Numerical ($P^N$) and analytical ($P^A$) posterior probabilities of the 1$\sigma$ intervals around \vtanI\ and \vtanII\ 
estimated using numerical and best fit posterior distribution function.}
\label{tab:vtanA2}
\begin{tabular}{ccccc}
\hline
Sample & $P_{I}^A$ & $P_{II}^A$ & $P_I^N$ & $P_{II}^N$ \\
\hline
Pr.1, CS &    0.16 &    0.27 &    0.13 &    0.37 \\
Pr.1, Rand &    0.23 &    0.12 &    0.18 &    0.16 \\
\hline
Pr.2, CS &    0.15 &    0.28 &    0.15 &    0.32 \\
Pr.2, Rand &    0.34 &    0.16 &    0.30 &    0.17 \\
\hline
Pr.3, CS &    0.16 &    0.27 &    0.13 &    0.37 \\
Pr.3, Rand &    0.23 &    0.12 &    0.18 &    0.16 \\
\hline
Pr.4, CS &    0.15 &    0.28 &    0.15 &    0.32 \\
Pr.4, Rand &    0.35 &    0.16 &    0.30 &    0.17 \\
\hline
Pr.5, CS &    0.17 &    0.23 &    0.16 &    0.32 \\
Pr.5, Rand &    0.30 &    0.13 &    0.21 &    0.21 \\
\hline
\end{tabular} 
\end{center}
\end{table}

\label{lastpage}

\end{document}